\documentclass[a4paper,11pt]{article}
\usepackage[hmargin=3cm,vmargin=3cm]{geometry,}
\usepackage{graphicx, amssymb, textcomp, hyperref, amsmath}
\numberwithin{equation}{section}
\title{\bf Unified derivation of exact solutions for a class of quasi-exactly solvable models}
\author{\Large Davids Agboola \footnote{d.agboola@maths.uq.edu.au}~ and~ Yao-Zhong Zhang \footnote{yzz@maths.uq.edu.au}}
\date{\it School of Mathematics and Physics, The University of Queensland, \\Brisbane, QLD 4072, Australia}
\begin{document}
\maketitle
\vspace{0.5in}
\noindent {\bf Abstract:} We present a unified treatment of exact solutions for a class of four
quantum mechanical models, namely the anharmonic singular potential,
the generalized quantum isotonic oscillator, 
the soft-core Coulomb potential, and the non-polynomially modified oscillator. 
We show that all four cases are reducible to the same basic ordinary differential equation, 
which is quasi-exactly solvable.
A systematic and closed form solution to the basic equation is obtained via the Bethe ansatz method. 
Using the result, general exact expressions for the energies and the allowed potential parameters 
are given explicitly for each of the four cases in terms of the roots of a set of algebraic equations. 
A hidden $sl(2)$ algebraic structure is also discovered in these models.

\vspace{.3in}
\noindent{\em PACS numbers:} 03.65.-w, 03.65.Fd, 03.65.Ge, 02.30.Ik

\noindent{\em Keywords:} Quasi-exactly solvable systems, Bethe ansatz

\vspace{0.5in}
\section{Introduction}
Over the years, much efforts have been devoted to the determination of closed form solutions 
to (Schr\"odinger) differential equations of the form $H\Psi=E\Psi$, where $H$ is the Hamiltonian and $E$ the eigenvalue. 
In certain cases, the concerned equation can be reduced via suitable substitutions and 
transformations to a well known differential equation, from which solutions to the original problem
 can be easily obtained. For instance, closed form solutions to many exactly solvable models in quantum mechanics 
have been studied in connection with some well known classical differential equations. 
However, due to limited applications of exactly solvable systems in quantum mechanics, 
recent attentions have been on systems with partially, algebraically  solvable spectra. 
Such systems are said to be quasi-exactly solvable. Thus a quantum mechanical system 
is called  quasi-exactly solvable if only a finite number of eigenvalues and 
corresponding eigenvectors can be obtained exactly through algebraic means \cite{Turbiner96,GKO93,Ushveridze94}.  

An essential feature of a quasi-exactly solvable system is that the coefficients of the power series solutions 
to the underlying differential equations satisfy three- or more step recursion relations,
in contrast to the two-step recursions for the exactly solvable cases. The complexity of
the three- or more step recursion relations makes it very hard (if not impossible) to get exact power series solutions of
such systems. However, one can terminate the series at certain power of the variable 
by imposing certain constraints on the system parameters. 
By so doing, exact (polynomial) solutions to the system can be obtained, but only for certain energies 
and for special values of the parameters of the problem.

Solutions to quasi-exactly solvable systems have mostly been discussed in terms of the recursion 
relations of the power series coefficients. However, this approach does not generally allow one 
to give explicit, closed form algebraic expressions for the allowed potential parameters 
of the systems. In this paper, we give a systematic and unified algebraic 
treatment to a class of four quantum mechanical systems, namely, 
\begin{eqnarray*}
&&{\rm 1.~Singular~ anharmonic~ potential~ \cite{Znojil90,Kaushal91}:}~~ V(r)=\frac{1}{2}\omega^2 r^2+\frac{e}{r^4}+\frac{d}{r^6},\\
&&{\rm 2.~Generalized~ quantum~ isotonic~ oscillator~ \cite{CPRS08}:}~~ V(r)=\frac{1}{2}\omega^2 r^2+g\frac{r^2-a^2}{(r^2+a^2)^2},\\ 
&&{\rm 3.~Soft~ core~ Coulomb~ potential~ \cite{Znojil83}:}~~ V(r)=\frac{G}{r}-\frac{F}{r+\beta}, \\
&&{\rm 4.~Non~ polynomially~ modified~  oscillator~ \cite{RV67}-\cite{MP85}:}~~ V(r)=\frac{1}{2}\omega^2 r^2+\frac{\lambda r^2}{1+\delta r^2}. 
\end{eqnarray*}
Model 1 was investigated in \cite{Znojil90} using the Laurent series ansatz and continued fractions,
and its ground state was found in \cite{Kaushal91}. Solutions to models 2, 3 and 4 were studied in 
\cite{Sesma10}-\cite{PM91} on a case by case basis by means of the recursion relations.
We will show that all four models are reducible to the same basic differential equation,
which is quasi-exactly solvable. We solve these models exactly 
by using the functional Bethe ansatz method in \cite{Zhang11}. 
Our method allows us to obtain the explicit, closed form expressions of the energies and 
the allowed potential parameters for all cases once and for all in terms of the roots of the algebraic 
(Bethe ansatz) equations. Particularly interesting are our exact results for model 1, in that to our knowledge
this model was not previously recognized to be quasi-exactly solvable.   
We also find a underlying $sl(2)$ algebraic structure in all these models, 
which is responsible for the quasi-exact solvability.

The work is  organized as follows. In section 2, we define the four models and transform the corresponding 
differential equations into the same basic form. In section 3, we present exact polynomial 
solutions to the basic equation underlying  the four cases by using the Bethe ansatz method. 
We then give the general, closed form expressions of both the energies and the allowed 
parameters for each of the four cases. 
Hidden $sl(2)$ algebraic structures underlying the four models
are presented in section 4.  In section 5 we provide some concluding remarks.

\section{The four models and the underlying differential equations}
In this section, we give a description of the four models to be considered in this paper. 
For each case, we reduce the underlying differential equation to the basic equation which is quasi-exactly solvable.

\subsection{Singular anharmonic potential}
We consider a quantum mechanical model with the singular anharmonic potential \cite{Znojil90,Kaushal91},
\begin{equation}\label{anharmonic-V}
V(r)=\frac{1}{2}\omega^2 r^2+\frac{e}{r^4}+\frac{d}{r^6},
\end{equation}
where $\omega, e, d>0$ are constant parameters.  The corresponding radial Schr\"odinger equation  is given by
\begin{equation}\label{anharmonic-Eq1}
\frac{d^2 Y(r)}{dr^2}-\left[\frac{\ell(\ell+1)}{r^2}+\omega^2r^2+\frac{2e}{r^4}+\frac{2d}{r^6}\right] Y(r)+2E\, Y(r)=0, 
\end{equation}
where $\ell=-1, 0, 1,\cdots$, and  $E$ is the energy eigenvalue. 

Making the substitution,
\begin{equation}\label{anharmonic-Trans}
Y(r)=r^{3/2+e/\sqrt{2d}}\,\exp\left[-\frac{\omega}{2}r^2-\frac{\sqrt{2d}}{2}\frac{1}{r^2}\right]y(r),
\end{equation}
we obtain
\begin{equation}\label{anharmonic-Eq2}
y''(r)+\frac{2}{r}\left(-\omega r^2+\frac{3}{2}+\frac{e}{\sqrt{2d}}+\frac{\sqrt{2d}}{r^2}\right)y'(r)
+2\left[E-\omega\left(2+\frac{e}{\sqrt{2d}}\right)\right]y(r)
$$
$$
=\frac{1}{r^2}\left[2\omega \sqrt{2d}+\left(\ell+\frac{1}{2}\right)^2-\left(\frac{e}{\sqrt{2d}}+1\right)^2\right]y(r)
\end{equation}
Then the change of variable $t=r^2$ transforms the above differential equation into the form,
\begin{equation}\label{anharmonic-Eq3}
t^2y''(t)+\left[-\omega t^2+\left(2+\frac{e}{\sqrt{2d}}\right)t+\sqrt{2d}\right]y'(t)
+\frac{t}{2}\left[E-\omega\left(2+\frac{e}{\sqrt{2d}}\right)\right]y(t)
$$
$$
=\frac{1}{4}\left[2\omega \sqrt{2d}+\left(\ell+\frac{1}{2}\right)^2-\left(\frac{e}{\sqrt{2d}}+1\right)^2\right]y(t).
\end{equation}

\subsection{Generalized quantum isotonic oscillator}
The generalized quantum isotonic oscillator has recently attracted a number interests. 
This system is interesting because it is endowed with properties 
closely related to those of the harmonic oscillator.  
The radial Schr\"odinger equation  reads
\begin{equation}\label{eq:1}
\frac{d^2\Psi(r)}{dr^2}-\left[\frac{\ell(\ell+1)}{r^2}+\omega^2r^2
+2g\frac{r^2-a^2}{(r^2+a^2)^2}\right]\Psi(r)+2E\Psi(r)=0, 
\end{equation}
where $\omega$, $g>0$ and $a$ are constant parameters of the systems; $\ell=-1, 0, 1,\cdots$, 
and  $E$ is the energy eigenvalue. Making the variable change $z=\omega r^2$, \eqref{eq:1} becomes
\begin{equation}\label{eq:2}
z\frac{d^2\Psi(z)}{dz^2}+\frac{1}{2}\frac{d\Psi(z)}{dz^2}-\left[\frac{z}{4}
+\frac{\ell(\ell+1)}{4z}+\frac{g\left(z-\omega a^2\right)}{2\left(z+\omega a^2\right)^2}\right]\Psi(z)
=-\frac{E}{2\omega}\Psi(z)
\end{equation}
This equation can be transformed, by the substitution,
\begin{equation}\label{eq:3}
\Psi(z)=(z+\omega  a^2)^{b+1}~z^{\frac{\ell+1}{2}}e^{-z/2}~\psi(z),
\end{equation}
into the form
$$z\psi ''(z)+\left[\frac{7}{2}+2b+\ell-\frac{2(b+1)\omega a^2}{z+\omega a^2}-z\right]
\psi '(z)+\left[\frac{g/2+(b+1)(\ell+\omega a^2+3/2)}{z+\omega a^2}\right]\psi(z)$$
\begin{equation}\label{eq:4}
\hspace{2in}=\left[-\frac{E}{2\omega}+b+\frac{\ell}{2}+\frac{7}{4}\right]\psi(z),
\end{equation}
where 
\begin{equation}\label{eq:5}
g=b(b+1)\hspace{.3in}\Rightarrow\hspace{.3in}b=b_-=-\frac{1}{2}-\frac{1}{2}\sqrt{4g+1}. 
\end{equation}
We find it convenient to work with the new variable $t=z+\omega a^2$. Then Eq.\,\eqref{eq:4} becomes 
\begin{equation}\label{eq:7}
t(t-\omega a^2)\psi ''(t)+\left[-t^2+\left(\frac{5}{2}-\sqrt{4g+1}+\ell+\omega a^2\right)t
+\omega a^2\left(\sqrt{4g+1}-1\right)\right]\psi '(t)\hspace{0.2in}$$
$$+\frac{t}{2}\left(\frac{E}{\omega}+\sqrt{4g+1}-\ell-\frac{5}{2}\right)\psi(t)
=\left[-\frac{g}{2}+\frac{1}{2}\left(\sqrt{4g+1}-1\right)\left(\ell+\omega a^2+\frac{3}{2}\right)\right]\psi(t). 
\end{equation}

\subsection{Soft-core Coulomb potential}
We now consider the soft-core Coulomb potential \cite{Znojil83}
\begin{equation}\label{eq:9}
V(r)=\frac{G}{r}-\frac{Z}{r+\beta},
\end{equation}
where $G\neq Z>0$ and $\beta>0$ are constant parameters. Such potential is of interest 
in atomic and molecular physics. The $G=0$ case simulates the field of 
a smeared charge and is useful in describing mesonic atoms.  Solutions to the $G=0$ case has been studied in 
\cite{HSS10} using the method of solving the recursive relations. The parameter $\beta $ 
can be related to the strength of a laser field, within the range of $\beta=20$ - $40$ 
covering the experimental laser field strength \cite{LM81}. 

The corresponding reduced Schr\"odinger equation  is
\begin{equation}\label{eq:10}
\frac{d^2\Phi(r)}{dr^2}-\left[\frac{\ell(\ell+1)}{r^2}+\frac{2G}{r}-\frac{2Z}{r+\beta}\right]\Phi(r)+2E\Phi(r)=0,
\end{equation}
where $E<0$ is the energy eigenvalue and $\ell=-1,0,1,\cdots$. Making the substitution
\begin{equation}\label{eq:11}
\Phi(r)=(r+\beta)\,r^{\ell+1}e^{-c(r+\beta)}\phi(r),
\end{equation}
we have 
\begin{equation}\label{eq:12}
\phi''(r)+2\left[\frac{\ell+1}{r}+\frac{1}{r+\beta}-c\right]\phi '(r)+2\left[-\frac{(\ell+1)(c-1/\beta)+G}{r}
+\frac{Z-c-(\ell+1)/\beta}{r+\beta}\right]\phi(r)
$$
$$=-(2E+c^2)\phi(r).
\end{equation}
Setting 
\begin{equation}
E=-\frac{1}{2}c^2,
\end{equation}
we obtain 
\begin{equation}\label{eq:13}
t(t+\beta)\phi''(t)+2\left[-ct^2+(-c\beta+\ell+2)t+(\ell+2)\beta\right]\phi'(t)+2t\left[Z-G-c(\ell+2)\right]\phi(t)
$$
$$ =2\left[(\ell+1)\beta c+\beta G-\ell-1\right]\phi(t),
\end{equation}
where we have introduced the variable $t\equiv r$ for later convenience.

\subsection{Non-polynomially modified  oscillator}
The non-polynomially modified  oscillator potential 
has been found to be useful in several aspects of physics.  In laser physics, it arises out of the Fokker-Planck equation 
for a single-mode laser \cite{RV67}. In field theory it provides a simple zero-dimensional 
model possessing a non-polynomial Lagrangian \cite{BDSSV73}. The $\delta>0$ case is useful in toroidal plasma \cite{CS83}. 

The corresponding radial Schr\"odinger equation reads  
\begin{equation}\label{eq:17}
\frac{d^2\Xi(r)}{dr^2}-\left[\frac{\ell(\ell+1)}{r^2}+\omega^2r^2+2\lambda\frac{ r^2}{1+\delta r^2}\right]\Xi(r)+2E\Xi(r)=0 
\end{equation}
where $\omega, \delta>0$ and $\lambda$ and constant parameters; $\ell=-1,0,1,\cdots,$ and $E$ is the energy. 
With the variable change $z=\omega r^2$, Eq.(\ref{eq:17}) becomes
\begin{equation}\label{eq:17b}
z\frac{d^2\Xi(z)}{dz^2}+\frac{1}{2}\frac{d\Xi(z)}{dz}
-\frac{1}{4}\left[z+\frac{\ell(\ell+1)}{z}+\frac{2\lambda}{\omega}\frac{z}{\omega+\delta z}\right]\Xi(z)+\frac{E}{2\omega}\Xi(z)=0 
\end{equation}
Making the transformation,
\begin{equation}\label{eq:18}
\Xi(z)=\left(z+\frac{\omega}{\delta}\right)z^{\frac{\ell+1}{2}}\exp\left[-\frac{1}{2}z\right]\xi(z), 
\end{equation}
Eq.\,\eqref{eq:17b} becomes
\begin{equation}\label{eq:19}
z\xi ''(z)+\left(-z+\frac{2z}{z+\omega/\delta}+\ell+\frac{3}{2}\right)\xi '(x)
+\frac{\lambda/2\delta^2+\omega/\delta+\ell+\frac{3}{2}}{z+\omega/\delta}\xi(z)
$$
$$=\frac{1}{2}\left[-\frac{E}{\omega}+\frac{\lambda}{\omega\delta}+\ell+\frac{7}{2}\right]\xi(z).
\end{equation}
It is more convenient to work with the new variable $t=z+\omega/\delta$. Then Eq.\,\eqref{eq:19} becomes
\begin{equation}\label{eq:20}
t\left(t-\frac{\omega}{\delta}\right)\xi ''(t)+\left[- t^2+\left(\frac{\omega}{\delta}+\ell+\frac{7}{2}\right)t
-\frac{2\omega}{\delta}\right]\xi '(t)
+\frac{t}{2}\left[\frac{E}{\omega}-\frac{\lambda}{\omega\delta}-\ell-\frac{7}{2}\right]\xi(t)
$$
$$=-\left[\frac{\lambda}{4\delta^2}+\frac{\omega}{\delta}+\ell+\frac{3}{2}\right]\xi(t).
\end{equation}

\section{The basic equation and solutions to the four models}
It was demonstrated in the last section that the Schr\"odinger equations for the four cases can be transformed to 
Eqs.\,\eqref{anharmonic-Eq3}, \eqref{eq:7}, \eqref{eq:13} and \eqref{eq:20}, respectively, after the appropriate substitutions 
and variable changes. These four equations have the same basic form,
\begin{equation}\label{eq:3.1}
t(t-\alpha)\,S''(t)+\left[b_2t^2+b_1t+b_0\right]\,S'(t)+c_1\,t\,S(t)=c_0\,S(t),
\end{equation}
where $\alpha, b_2, b_1, b_0, c_1, c_0$ are constants. 
This equation has regular singular points $t=0, \alpha$ and confluently irregular singular point $t=\infty$. 
It is related, via a simple transformation, to the so-called generalized spheroidal wave equation (GSWE) 
(see e.g. \cite{Leaver86,Liu92}),
\begin{equation}\label{GSWE1}
t(t-t_0) X''(t)+\left(B_1+B_2\,t\right)X'(t)+\left[\Omega^2t(t-t_0)-2 k \Omega(t-t_0)+B_3\right]X(t)=0,
\end{equation}
where $t_0, B_1, B_2, B_3, k, \Omega$ are (possibly complex) constants.  Indeed, making the substitution
 $X(t)=e^{i\Omega t}S(t)$ in the generalized spheroidal wave equation  yields
\begin{equation}\label{GSWE2} 
t(1-t_0)S''(t)+\left[2i\Omega t^2+(B_2-2i\Omega t_0)t+B_1\right]\,S'(t)+\Omega(iB_2-2k)t\,S(t)
$$
$$
=-\left(i\Omega B_1+2 k\Omega t_0+B_3\right)\,S(t).
\end{equation}
This is nothing but the basic equation \eqref{eq:3.1} with the identifications,
\begin{equation}\label{identifications}
t_0=\alpha,~~~~\Omega=-\frac{i}{2}b_2,~~~~k=\frac{i}{2}\left(b_1+\alpha b_2-\frac{2c_1}{b_2}\right),
$$
$$
B_1=b_0,~~~~B_2=b_1+\alpha b_2,~~~~
B_3=\alpha c_1-c_0-\frac{b_2}{2}\left(b_0+\alpha b_1+\alpha^2 b_2\right).
\end{equation}
Note that for the four models considered in this paper, the constants $\alpha, b_2, b_1, b_0, c_1, c_0$ 
in the basic equation are real numbers. 

The basic equation (or more generally the GSWE) is quasi-exactly solvable for 
certain values of its parameters, and exact solutions are 
given by degree $n$ polynomials in $t$ with $n$ being non-negative integers. 
In fact, the basic equation is a special case of the general 2nd order differential equations
solved in \cite{Zhang11} by means of the Bethe ansatz method. Applying the results in \cite{Zhang11}, we have  
\\\\
{\bf Proposition 3.1} Let $n$ be any non-negative integer.
The basic differential equation Eq.\,\eqref{eq:3.1} has degree $n$ polynomial solutions 
\begin{equation}\label{eq:3.3}
S(t)=\prod_{i=1}^n(t-t_i),\hspace{0.2in}S(t)\equiv 1\hspace{0.1in}\mbox{for}\hspace{0.1in} n=0
\end{equation}
with distinct roots $t_1, t_2,\dots,t_n$  only if the constant coefficients in \eqref{eq:3.1} satisfy the constraints
\begin{equation}\label{eq:3.5}
c_1=-nb_2,
\end{equation}
\begin{equation}\label{eq:3.6}
c_0=n(n-1)+b_2\sum_{i=1}^nt_i+nb_1,
\end{equation}
where the roots $t_1,t_2,\dots,t_n$ are determined by the Bethe ansatz equations
\begin{equation}\label{eq:3.7}
\sum_{j\neq i}^n\frac{2}{t_i-t_j}=-\frac{b_2t_i^2+b_1t_i+b_0}{t_i(t_i-\alpha)},\hspace{0.3in} i=1,2,\dots,n.
\end{equation}

In the following, we apply the above results to derive general, explicit and closed form expressions
for the energies and the allowed potential parameters of the four models. 

\subsection{Singular anharmonic potential}
In this case, $S(t)=y(t)$, $\alpha=0$, $b_2=-\omega$, $b_1=2+\frac{e}{\sqrt{2d}}$, $b_0=\sqrt{2d}$, 
$c_1=\frac{1}{2}\left[E-\omega\left(2+\frac{e}{\sqrt{2d}}\right)\right]$ and 
$c_0=\frac{1}{4}\left[2\omega \sqrt{2d}+\left(\ell+\frac{1}{2}\right)^2-\left(\frac{e}{\sqrt{2d}}+1\right)^2\right]$.
Then by Eqs. \eqref{eq:3.5} and \eqref{eq:3.6}, we obtain the closed form expressions of the energies
and wave functions
\begin{equation}\label{anharmonic-E}
E_n=\omega\left(2n+2+\frac{e}{\sqrt{2d}}\right)$$$$
Y_n(r)=r^{3/2+e/\sqrt{2d}}\left[\prod_{i=1}^n(r^2-t_i)\right]\,\exp\left[-\frac{\omega}{2}r^2-\frac{\sqrt{2d}}{2}\frac{1}{r^2}\right]
\end{equation} 
and the constraint for the potential parameters
\begin{equation}\label{anharmonic-Constraint}
2\omega\left(\sqrt{2d}+2\sum_{i=1}^nt_i\right)+\left(\ell+\frac{1}{2}\right)^2
=4n\left(n+1+\frac{e}{\sqrt{2d}}\right)+\left(\frac{e}{\sqrt{2d}}+1\right)^2,
\end{equation}
where the roots $\{t_i\}$ are determined by the Bethe ansatz equations,
\begin{equation}\label{anharmonic-BAEs}
\sum_{j\neq i}^n\frac{2}{t_i-t_j}=\frac{\omega t_i^2-\left(2+\frac{e}{\sqrt{2d}}\right)t_i-\sqrt{2d}}{t_i^2},~~~~i=1,2,\cdots,n.
\end{equation}

It is easy to see that $y=1$ is a solution of \eqref{anharmonic-Eq3} provided the potential parameters satisfy a constraint.
This solution corresponds to the $n=0$ case in the general expressions above. Indeed, from Eqs. \eqref{anharmonic-E}
and \eqref{anharmonic-Constraint}, we obtain 
\begin{equation}\label{anharmonic-E0}
E_0=\omega\left(2+\frac{e}{\sqrt{2d}}\right)
$$$$
Y_0(r) = r^{3/2+e/\sqrt{2d}}\,\exp\left[-\frac{\omega}{2}r^2-\frac{\sqrt{2d}}{2}\frac{1}{r^2}\right].
\end{equation} 
and the constraint
\begin{equation}\label{anharmonic-Constraint0}
2\omega\sqrt{2d}+\left(\ell+\frac{1}{2}\right)^2=\left(\frac{e}{\sqrt{2d}}+1\right)^2.
\end{equation}
This wave function has no nodes and so the state described by it is the ground state of the system. Eqs.\,
\eqref{anharmonic-E0} and \eqref{anharmonic-Constraint0} reproduce the ground state solution
and the corresponding constraint for the parameters found in \cite{Kaushal91}.

For $n=1$, Eqs. \eqref{anharmonic-E}, \eqref{anharmonic-Constraint} and \eqref{anharmonic-BAEs} give rise to the energy
\begin{equation}\label{anharmonic-E1}
E_1=\omega\left(4+\frac{e}{\sqrt{2d}}\right)
\end{equation} 
and the constraint for the potential parameters
\begin{equation}\label{anharmonic-Constraint1}
2\omega\left(\sqrt{2d}+2t_1\right)+\left(\ell+\frac{1}{2}\right)^2=8+\frac{2e}{\sqrt{2d}}+\left(\frac{e}{\sqrt{2d}}+1\right)^2,
\end{equation}
where the root $t_1$ is determined by the Bethe ansatz equation,
\begin{equation}\label{anharmonic-BAEs1}
\omega t_1^2-\left(2+\frac{e}{\sqrt{2d}}\right)t_1-\sqrt{2d}=0
$$
$$\Rightarrow\hspace{0.1in}t_1=\frac{1}{2\omega}\left(2+\frac{e}{\sqrt{2d}}\pm\sqrt{\left(2+\frac{e}{\sqrt{2d}}\right)^2
   +4\omega\sqrt{2d}}\right).
\end{equation}
It follows that the potential parameters obey 
\begin{equation}\label{anharmonic-Constraint1b}
\frac{1}{4}\left[\frac{e^2}{2d}-2\omega\sqrt{2d}+5+\frac{4e}{\sqrt{2d}}-\left(\ell+\frac{1}{2}\right)^2\right]^2
={\left(2+\frac{e}{\sqrt{2d}}\right)^2+4\omega\sqrt{2d}}.
\end{equation}
The corresponding wave function is
\begin{equation}\label{anharmonic-Wavefunction1}
Y_1(r) = r^{3/2+e/\sqrt{2d}}\,\left(r^2-t_1\right)\,\exp\left[-\frac{\omega}{2}r^2-\frac{\sqrt{2d}}{2}\frac{1}{r^2}\right]
\end{equation}
with $t_1$ given by \eqref{anharmonic-BAEs1} and $\omega, e, d$ constrained by \eqref{anharmonic-Constraint1b}. 
This wave function gives rise to the first excited state of the system.

\subsection{Generalized quantum isotonic oscillator}
In this case,  $S(t)=\psi(t)$,   $\alpha=\omega a^2$, $b_2=-1$, $b_1=\frac{5}{2}-\sqrt{4g+1}+\ell+\omega a^2$, 
$b_0=\omega a^2(\sqrt{4g+1}-1)$, $c_1=\frac{1}{2}\left(\frac{E}{\omega}+\sqrt{4g+1}-{\ell}-\frac{5}{2}\right)$ 
and $c_0=-g/2+\frac{1}{2}(\sqrt{4g+1}-1)(\ell+\omega a^2+3/2)$. 
Here $b$ is given in (\ref{eq:5}).
Then by Eqs. \eqref{eq:3.5} and \eqref{eq:3.6}, we obtain the general, closed form expressions of the energies
and wave functions  
\begin{equation}\label{eq:4.1}
E_n=\omega\left(2n+{\ell}+\frac{5}{2}-\sqrt{4g+1}\right),
$$$$
\Psi_n(r)\sim(r^2+a^2)^{b+1}~r^{{\ell+1}}e^{-\omega r^2/2}~\left[\prod_{i=1}^n\left( r^2+ a^2-\frac{t_i}{\omega}\right)\right],
\end{equation}
and the constraint for the potential parameters,
\begin{equation}\label{eq:4.2}
n\left(n+\frac{3}{2}-\sqrt{4g+1}+\ell+\omega a^2\right)-\sum_{i=1}^nt_i
$$
$$=-\frac{g}{2}+\frac{1}{2}(\sqrt{4g+1}-1)(\ell+\omega a^2+3/2) 
\end{equation}
where the roots $t_1,t_2,\dots t_n$ obey the Bethe ansatz equations 
\begin{equation}\label{eq:4.3}
\sum_{j\neq i}^n\frac{2}{t_i-t_j}=\frac{t_i^2-\left(\ell+\frac{5}{2}-\sqrt{4g+1}+\omega a^2\right)t_i
+\omega a^2(1-\sqrt{4g+1})}{t_i(t_i-\omega a^2)},
$$
$$i=1,2,\cdots n.
\end{equation}

It is easily seen that $\psi=1$ is a solution of \eqref{eq:7} provided that a constraint is satisfied by the parameters.
indeed, this solution corresponds to the $n=0$ case in the general expressions above.
{} From Eqs.\,\eqref{eq:4.1} and \eqref{eq:4.2}, we have 
\begin{equation}\label{eq:4.4}
E_0=\omega\left(\ell+\frac{5}{2}-\sqrt{4g+1}\right),
$$$$
\Psi_0(r)\sim(r^2+a^2)^{b+1}~r^{{\ell+1}}e^{-\omega r^2/2},
\end{equation}
and 
\begin{equation}\label{eq:4.5}
g+\ell+\frac{3}{2}+\omega a^2=\left(\ell+\frac{3}{2}+\omega a^2\right)\sqrt{4g+1}  
$$$$\Rightarrow~~~~
g=2\left(\ell+1+\omega a^2\right)\left(2\ell+3+2\omega a^2\right). 
\end{equation}
Hence the energy and wave function of the system can be written as  
\begin{equation}\label{eq:4.8}
E_0= -\omega\left(\frac{5}{2}+3\ell+4 \omega a^2\right) 
\end{equation}
and 
\begin{equation}\label{eq:4.9}
\Psi_0(r) \sim r^{\ell+1}\left(r^2+a^2\right)^{-2(\ell+1+\omega a^2)}e^{-\frac{1}{2}\omega r^2}.
\end{equation}
The wave functions $\Psi_0(r)$ does not have nodes and so the state described by this is the ground states. 

For $n=1$, Eqs.\,\eqref{eq:4.1}, \eqref{eq:4.2} and \eqref{eq:4.3} give
\begin{equation}\label{eq:4.14}
E_1=\omega\left(\ell+\frac{9}{2}-\sqrt{4g+1}\right),
\end{equation}
\begin{equation}\label{eq:4.15}
g+\frac{13}{2}+3\left(\ell+\omega a^2\right)=2t_1+\left(\ell+\omega a^2+\frac{7}{2}\right)\sqrt{4g+1}
\end{equation}
and
\begin{equation}\label{eq:4.16}
t_1=\frac{1}{2}\left(2s_1+s_2+\omega a^2\pm\sqrt{(2s_1+s_2)^2+\omega a^2(2s_2-4s_1+\omega a^2)}\right),
\end{equation}
where $s_1=\frac{1}{2}\left(1-\sqrt{4g+1}\right)$ and $s_2=\ell+3/2$.
Substituting Eq.\,\eqref{eq:4.16} into Eq.\,\eqref{eq:4.15}, we have the following constraint 
for the allowed parameter $g$,
\begin{equation}\label{eq:4.17}
g+2(\ell+\omega a^2+2)-\left(\frac{5}{2}+\ell+\omega a^2\right)\sqrt{4g+1}\hspace{1in}$$
$$\hspace{1.5in}=\pm\sqrt{\left(\frac{5}{2}+\ell-\sqrt{4g+1}\right)^2+\omega a^2\left(2\ell+\omega a^2 +1 +2\sqrt{4g+1}\right)},
\end{equation}
which can be solved to give
\begin{equation}\label{eq:4.18}
g=-\frac{1}{4}+\frac{1}{36}\left(2A^\frac{1}{3}+\frac{B}{A^{\frac{1}{3}}}+C\right)^2
\end{equation}
where 
\begin{eqnarray*}
A&=&-15\ell+93\omega a^2-30\ell^2-60\ell\omega a^2-30(\omega a^2)^2\nonumber\\
& &-24\ell^2\omega a^2-24\ell(\omega a^2)^2-8\ell^3-8(\omega a^2)^3+53\nonumber\\
& &+3\left[-1380\ell-108\omega a^2-1443\ell^2-3246\ell\omega a^2-507(\omega a^2)^2\right.\nonumber\\
& &-2556\ell^2\omega a^2-3276\ell(\omega a^2)^2-624\ell^3\omega a^2-1224(\ell\omega a^2)^2\nonumber\\
& &\left.-1008\ell(\omega a^2)^2-612\ell^3-1332(\omega a^2)^3-108\ell^4-300(\omega a^2)^4-450\right]^\frac{1}{2},\nonumber\\
B&=&38+8\ell^2+16\ell\omega a^2+20\ell+8(\omega a^2)^2+20\omega a^2,\nonumber\\
C&=&8\omega a^2+8\ell+19.
\end{eqnarray*}
Hence we obtain the energy and wave function  
\begin{equation}\label{eq:4.19}
E_1=\omega\left[\ell+\frac{9}{2}-\frac{1}{3}\left(2A^\frac{1}{3}+\frac{B}{A^{\frac{1}{3}}}+C\right)\right],
$$$$
\Psi_1(r)\sim r^{\ell+1}\left(r^2+a^2\right)^{b+1}e^{-\frac{1}{2}\omega r^2}\left(r^2+a^2-\frac{t_1}{\omega}\right)
\end{equation}
where
\begin{equation}\label{eq:4.21}
t_1=\frac{1}{2}\left[\ell+\omega a^2+\frac{5}{2}-x\pm\sqrt{\left(\ell+\frac{5}{2}-x\right)^2
+\omega a^2\left(2\ell+1+\omega a^2+2x\right)}\right]
\end{equation}
with
\begin{equation}\label{eq:22}
x=\frac{1}{3}\left(2A^{\frac{1}{3}}+\frac{B}{A^{\frac{1}{3}}}+C\right).
\end{equation}

\subsection{Soft-core Coulomb potential}
In this case,  $S(t)=\phi(t)$,  $\alpha=-\beta$, $b_2=-2c$, $b_1=2(-\beta c+\ell+2)$, $b_0=(\ell+1)\beta$, 
$c_1=2[Z-G-(\ell+2)c]$ and $c_0=2[\beta (\ell+1)c + \beta G-\ell-1]$. 
Then by Eqs.\,\eqref{eq:3.5} and \eqref{eq:3.6}, we have for the energies, wavfunction
and the constraint for the parameter $Z$,
\begin{equation}\label{eq:4.32}
c=\frac{Z-G}{n+\ell+2}\hspace{0.1in}\Rightarrow\hspace{0.1in}E_{n}=-\frac{1}{2}\left(\frac{Z-G}{n+\ell+2}\right)^2,$$$$
\Phi_n(r)=(r+\beta)\,r^{\ell+1}e^{-c(r+\beta)}\left[\prod_{i=1}^n(r-t_i)\right],
\end{equation}
\begin{equation}\label{eq:4.33}
n\left(\frac{n+3}{2}+\ell-\beta c\right)-c\sum_{i=1}^nt_i=\beta (\ell+1)c+\beta G-\ell-1,
\end{equation}
where $t_i$ satisfy the Bethe ansatz equations
\begin{equation}\label{eq:4.34}
\sum_{j\neq i}^n\frac{1}{t_i-t_j}=\frac{ct_i^2-(-\beta c+\ell+2)t_i-(\ell+1)\beta}{t_i(t_i+\beta)},~~~~i=1,2,\cdots,n.
\end{equation}
Note that $c\neq 0$ as $Z\neq G$. 

It is easily seen that $\phi=1$ is a solution to (\ref{eq:13}) provided $c$ and $Z$ satisfy certain constraints.
This solution corresponds to the $n=0$ case of our general expressions \eqref{eq:4.32}-\eqref{eq:4.34}, 
from which we obtain the solutions 
\begin{equation}\label{eq:4.35}
E_0=-\frac{1}{2}\left(\frac{Z-G}{\ell+2}\right)^2,$$$$
\Phi_0(r)= (r+\beta)r^{\ell+1}\exp\left[-\frac{\ell+1-\beta G}{(\ell+1)\beta}(r+\beta)\right].
\end{equation} 
and the allowed values for the parameter $Z$,
\begin{equation}\label{eq:4.36}
Z=\frac{\ell+2}{\beta}-\frac{G}{\ell+1}.
\end{equation}
This wave function does not have nodes and so the state described by it is the ground state of the system.

For $n=1$, \eqref{eq:4.32}-\eqref{eq:4.34} become
\begin{equation}\label{eq:4.38a}
c=\frac{Z-G}{\ell+3}\hspace{0.1in}\Rightarrow\hspace{0.1in}E_{1}=-\frac{1}{2}\left(\frac{Z-G}{\ell+3}\right)^2,
\end{equation}
\begin{equation}\label{eq:4.38b}
ct_1+\beta(\ell+2) c=2\ell+3-\beta G,
\end{equation}
where the root $t_1$ obeys 
\begin{equation}\label{eq:4.38c}
ct_1^2-(-\beta c+\ell+2)t_1-(\ell+1)\beta=0\hspace{0.1in}
$$
$$\Rightarrow\hspace{0.1in}t_1=\frac{1}{2c}\left(-\beta c+\ell+2\pm
\sqrt{\beta^2c^2+2\ell\beta c+(\ell+2)^2}\right).
\end{equation}
It follows that the parameter $Z$ satisfies the constraint
\begin{equation}\label{eq:4.41}
(\ell+3)^2(\beta Z-\ell-1)(\beta Z-2\ell-3)+(2+\nu)\beta(Z-G)(-3\beta Z+4\ell+6)+2\beta^2(Z-G)^2=0.
\end{equation}
The corresponding wave function  is
\begin{equation}\label{eq:4.43}
\Phi_1(r)=(r+\beta)(r-\tau_1) r^{\ell+1}\exp\left[-\frac{Z-G}{\ell+3}(r+\beta)\right],
\end{equation}
where
\begin{equation}
\tau_1=\frac{1}{2}\left[-\beta+\frac{(\ell+1)(\ell+3)}{Z-G}\pm\sqrt{\beta^2+\frac{2\ell(\ell+3)}{Z-G}+
\left(\frac{(\ell+2)(\ell+3)}{Z-G}\right)^2}\right]
\end{equation}
with $Z$ determined by \eqref{eq:4.41}. This wave function gives the first excited state of the system.

\subsection{Non-polynomially modified  oscillator}
In this case,  $S(t)=\xi(t)$,  $\alpha=\frac{\omega}{\delta}$, $b_2=-1$, $b_1=\frac{\omega}{\delta}+\ell+2\eta+\frac{7}{2}$,
$b_0=-\frac{2\omega}{\delta}$, 
$c_1=\frac{1}{2}\left[\frac{E}{\omega}-\frac{\lambda}{\omega\delta}-\ell-2\eta-\frac{7}{2}\right]$ and 
$-c_0=\frac{\lambda}{4\delta^2}+\frac{\omega}{\delta}+\ell+2\eta+\frac{3}{2}$. 
Then from Eqs.\,\eqref{eq:3.5} and \eqref{eq:3.6}, we obtain the energies and wave functions,
\begin{equation}\label{eq:4.57}
E_n=\frac{\lambda}{\delta}+\omega\left(2n+\ell+\frac{7}{2}\right),
$$$$
\Xi_n(r)\sim\left(1+\delta r^2\right)r^{\ell+1}e^{-\frac{\omega}{2}r^2}~\left[\prod_{i=1}^n\left(1+\delta r^2-\frac{\delta}{\omega}t_i\right)\right], 
\end{equation}
and the allowed values for the parameter $\lambda$,
\begin{equation}\label{eq:4.58}
\frac{\lambda}{4\delta^2}=-(n+1)\left(n+\frac{\omega}{\delta}+\ell+\frac{3}{2}\right)+\sum_{i=1}^nt_i, 
\end{equation}
where the roots $t_i$ satisfy the Bethe ansatz equations
\begin{equation}\label{eq:4.59}
\sum_{j\neq i}^n\frac{2}{t_i-t_j}=\frac{t_i^2-({\omega}/{\delta}+\ell+\frac{7}{2})t_i+{2\omega}/{\delta}}
{t_i(t_i-\omega/\delta)},~~~~i=1,2,\cdots,n.
\end{equation}

It is easily seen that $\xi=1$ is a solution of \eqref{eq:20} provided that the parameter $\lambda$ obeys
certain constraint. This solution corresponds to the $n=0$ case in the general expressions above. 
The energy and the wavefunction are
\begin{equation}
E_0=\frac{\lambda}{\delta}+\omega\left(\ell+\frac{7}{2}\right),
$$$$\Xi_0(r)\sim\left(1+\delta r^2\right)r^{\ell+1}e^{-\frac{\omega}{2}r^2}
\end{equation}
with the allowed values for $\lambda$ given by
\begin{equation}
\frac{\lambda}{4\delta^2}=-\frac{\omega}{\delta}-\ell-\frac{3}{2}.
\end{equation}
\noindent The wave functions $\Xi_0(r)$ does not have nodes and thus the states described
by this is the ground state of the system.

For $n=1$, we have
\begin{equation}
E_1=\frac{\lambda}{\delta}+\omega\left(\ell+\frac{11}{2}\right),
\end{equation}
\begin{equation}
\frac{\lambda}{4\delta^2}=t_1-2\left(\frac{\omega}{\delta}+\ell+\frac{5}{2}\right),
\end{equation}
where the root $t_1$ obeys
\begin{equation}\label{eq:const-t}
t_1^2-\left(\frac{\omega}{\delta}+\ell+\frac{7}{2}\right)t_1+\frac{2\omega}{\delta}=0
$$
$$ \Rightarrow\hspace{0.1in}t_1=\frac{1}{2}\left(\frac{\omega}{\delta}+\ell+\frac{7}{2}\pm
\sqrt{\frac{\omega^2}{\delta^2}+\frac{\omega}{\delta}\left(2\ell-1\right)+\left(\ell+\frac{7}{2}\right)^2}\right).
\end{equation}
The energy and the first excited state wave function  are therefore
\begin{equation}\label{eq:465a}
E_1=(2\ell+3)(\omega-4\delta)+4\delta\left(t_1-2\right)-\left(\ell+\frac{11}{2}\right)\omega,
$$$$
\Xi_1(r)\sim \left(1+\delta r^2\right) r^{\ell+1}e^{-\frac{\omega}{2}r^2}
\left(1+\delta r^2-\frac{\delta}{\omega}t_1\right),
\end{equation}
where $t_1$ is given in Eq.\,\eqref{eq:const-t}. 

\section{Hidden Lie algebraic structure}
The basic equation \eqref{eq:3.1} unifying the underlying differential equations of the four  
models considered in this paper possesses a hidden Lie $sl(2)$ algebraic structure which is 
responsible for its quasi-exact solvability. 

Let us first of all recall some well-known facts. Consider the differential operators,
\begin{equation}\label{eq:4.31}
J^-=\frac{d}{dt},\hspace{0.2in}J^+=t^2\frac{d}{dt}-nt,\hspace{0.2in}J^0=t\frac{d}{dt}-\frac{n}{2}.
\end{equation}  
These operators satisfy the $sl(2)$ commutation relations for any value of the parameter $n$.
If $n$ is a non-negative integer, then  \eqref{eq:4.31} provides a
$n+1$-dimensional irreducible  representation,  ${\cal P}_{n+1}(t)=<1, t, t^2,\cdots, t^n>$ of the $sl(2)$ algebra. 
 From this it is evident that any differential operator which is a polynomial of the $sl(2)$ generators
\eqref{eq:4.31} will have the space ${\cal P}_{n+1}$ as its finite-dimensional invariant subspace,
i.e. possesses $(n+1)$ eigenfunctions in the form of a polynomial in $t$ of degree $n$.  
This is the main idea behind quasi-exact solvability of a differential operator with 
a $sl(2)$ algebraization (i.e. a hidden $sl(2)$ algebraic structure) \cite{Turbiner96,GKO93}.

Now consider the basic equation \eqref{eq:3.1} and write it as the form
\begin{equation}\label{HS=cS}
H S(t)=c_0\, S(t),~~~~~~ H=t(t-\alpha)\,\frac{d^2}{dt^2}+\left[b_2t^2+b_1t+b_0\right]\,\frac{d}{dt}+c_1\,t.
\end{equation}
Then it can be checked that, if $c_1=-nb_2$ with $n$ being any non-negative integer, the differential operator
$H$ above can be written as
\begin{equation}\label{H=JJetc}
H=J^0J^0-\alpha J^0J^-+b_2J^++[n-1+b_1]J^0+\left(-\frac{n}{2}\alpha+b_0\right)J^-+\frac{n}{2}\left[\frac{n}{2}-1+b_1\right].
\end{equation}
Thus \eqref{H=JJetc} provides a $sl(2)$ algebraization for the differential operator $H$. 
In other words, the differential equation \eqref{HS=cS} (or equivalently the the basic equation \eqref{eq:3.1}) 
has an underlying $sl(2)$ algebraic structure \footnote{See e.g. \cite{Turbiner94,CH01} for a similar $sl(2)$ 
algebraic structure in the relative motion of two charged particles in an external oscillator 
potential and the system describing charged particle moving in Coulomb and magnetic fields.} 
and is quasi-exactly solvable if $c_1=-nb_2$,
which is exactly the relation \eqref{eq:3.5}.  For such values of $c_1$, exact solutions  of the basic equation 
are given by the degree $n$ polynomials \eqref{eq:3.3}, with the allowed values of the constant coefficients 
$b_1, c_0$ being determined by the constraint \eqref{eq:3.6} and the Bethe ansatz equations \eqref{eq:3.7}.

Translating these results back to the original four cases, we can conclude that
the four quantum mechanical models considered in this paper possess a hidden $sl(2)$ algebraic structure, 
which is responsible for their quasi-exact solvability.

\section{Concluding remarks}
We have provided a unified treatment of the four quasi-exactly solvable quantum mechanical models 
and shown that the corresponding radial Schr\"odinger equations are reducible to 
the same basic equation which can be exactly solved by using 
the Bethe ansatz method. For each of the four cases, we have derived the closed form expressions of the energies,
wave functions and the allowed potential parameters. We have also shown that
all four cases possess a hidden $sl(2)$ algebraic structure, which is responsible for the
quasi-exact solvability of the systems. Let us remark, however, that the existence of a underlying Lie algebraic 
structure in a differential equation is only a sufficient condition for the differential equation to be 
quasi-exactly solvable. In fact there are more general (than the Lie-algebraically based)
differential equations which do not possess a underlying Lie algebraic structure but are
nevertheless quasi-exactly solvable (i.e have exact polynomial solutions) \cite{Zhang11}. 
\footnote{A general 2nd order quasi-exactly solvable differential equation has a 
underlying Lie algebraic structure if the coefficients
of the 1st and 2nd order derivatives are algebraically dependent. See Corollary 1.3 and its proof of \cite{Zhang11}.} 

We have seen that the basic equation \eqref{eq:3.1} unifying the four cases considered in this paper 
is a GSWE. Thus as by-products of the results of this paper, the GSWE possesses a hidden $sl(2)$ algebraic symmetry 
and is quasi-exactly solvable. And its exact (polynomial) solutions are given by expressions 
\eqref{eq:3.3}-\eqref{eq:3.7} in proposition 3.1 together with the identifications \eqref{identifications}.
It is well-known that the GSWE has important applications in astrophysics and molecular physics.
For example, Teukolsky's equation governing perturbations of the Kerr black hole and the (radial or angular) equation describing
the scattering of a charged particle on two Coulomb centers with different charges (such as one-electron diatomic
molecules) are GSWE. It would be interesting to generalize the results of the present paper to obtain exact, 
closed form solutions to systems arising in the above-mentioned two specific physical contexts, e.g. the
one-electron diatomic molecular model. Research along this path is underway, and results will be reported elsewhere.

\section*{Acknowledgments} 
The second author gratefully acknowledges the support of the Australian Research Council. 
through Discovery Project DP110103434.

\end{document}